\documentclass[pra,twocolumn,longbibliography,superscriptaddress]{revtex4-1}
\usepackage[utf8]{inputenc}
\usepackage{amsmath}
\usepackage{amsfonts}
\usepackage{amssymb}
\usepackage{bm}
\usepackage{graphicx}
\usepackage{xcolor}
\usepackage{microtype}
\usepackage{ dsfont }

\begin{document}
\title{$\mathbb{Z}_2$ characterization for three-dimensional multiband Hubbard models}
\author{Bernhard Irsigler}
\affiliation{Institut f\"ur Theoretische Physik, Goethe-Universit\"at Frankfurt am Main, Germany}
\author{Jun-Hui Zheng}
\affiliation{Institut f\"ur Theoretische Physik, Goethe-Universit\"at Frankfurt am Main, Germany}
\affiliation{Center for Quantum Spintronics, Department of Physics,
Norwegian University of Science and Technology, NO-7491 Trondheim, Norway}
\author{Fabian Grusdt}
\affiliation{Munich Center for Quantum Science and Technology (MCQST), Schellingstraße 4, 80799 M\"unchen, Germany}
\affiliation{Fakult\"at f\"ur Physik, Ludwig-Maximilians-Universit\"at, 80799 M\"unchen, Germany}
\author{Walter Hofstetter}
\affiliation{Institut f\"ur Theoretische Physik, Goethe-Universit\"at Frankfurt am Main, Germany}
\begin{abstract}
We introduce three numerical methods for characterizing the topological phases of three-dimensional multiband Hubbard models based on twisted boundary conditions, Wilson loops, as well as the local topological marker. We focus on the half-filled, three-dimensional time-reversal-invariant Hofstadter model with finite spin-orbit coupling. Besides the weak and strong topological insulator phases we find a nodal line semimetal in the parameter regime between the two three-dimensional topological insulator phases. Using dynamical mean-field theory combined with the topological Hamiltonian approach we find stabilization of these three-dimensional topological states due to the Hubbard interaction. We study surface states which exhibit an asymmetry between left and right surface originating from the broken parity symmetry of the system. Our results set the stage for further research on inhomogeneous three-dimensional topological systems, proximity effects, topological Mott insulators, non-trivially linked nodal line semimetals and circuit-based quantum simulators.
\end{abstract}
\maketitle

\section{Introduction}
Three-dimensional (3d) topological states surpass their two-dimensional (2d) counterparts in terms of complexity and richness. A 2d quantum spin Hall (QSH) state, e.g., is characterized by a single $\mathbb{Z}_2$ number $\nu$, i.e., the system is either in a topologically trivial state $\nu=0$ or in the non-trivial QSH state $\nu=1$. The 3d analogue, however, is characterized by four $\mathbb{Z}_2$ numbers $(\nu_0;\nu_1,\nu_2,\nu_3)$ leading to a total of 16 topologically distinct states \cite{Fu2007,Moore2007,Roy2009}. The straightforward way of picturing this 3d generalization is by stacking many 2d QSH layers. If the coupling between these layers is weak one finds a weak topological insulator (WTI), e.g., $(\nu_0;\nu_1,\nu_2,\nu_3)=(0;0,0,1)$. A WTI exhibits robust, helical surface states \cite{Ringel2012,Sbierski2016} which encircle the $(\nu_1,\nu_2,\nu_3)$ axis. The strong topological insulator (STI), on the other hand, emerges if $\nu_0=1$ and features helical surface states in any direction. There is no spin-conserved backscattering of the surface states due to spin-momentum locking which is protected by the time-reversal invariance (TRI). 

A further difference to the 2d case is that in 3d also gapless topological states can emerge, such as Dirac and Weyl semimetals \cite{Murakami2008,Young2012,Armitage2018}. If nonsymmorphic symmetries are present, however, a symmetry-protected Dirac semimetal is predicted in 2d \cite{Young2015}. Another prominent example in 3d are nodal-line semimetals (NLSM) which exhibit a bulk band touching along a closed line embedded in the 3d Brillouin zone (BZ). These lines are not accidental but are topologically as well as symmetry protected and cannot simply gap out. A particle moving on a path in the 3d BZ linking the nodal line picks up a nontrivial Berry phase \cite{Fang2016}. Here, the nodal line acts as singularity around which the Berry phase is acquired. The only way to open a gap is to shrink the nodal line to a point which can then gap out. On the other hand, integrating the Berry curvature on a 2d manifold enclosing the complete nodal line can yield a nonzero Chern number. This corresponds to a topological charge similar to Weyl points in Weyl semimetals \cite{Burkov2011,Armitage2018,Hirayama2018}. If the nodal line carries a topological charge, it cannot gap out by simply shrinking to a point but has to recombine with another nodal line carrying the opposite topological charge \cite{Fang2016}. Even more complex physics occurs if one combines many nodal lines which are topologically non-trivially linked \cite{Bzdusek2016,Chang2017,Chen2017,Yan2017,Li2018a}.

\begin{figure}
\includegraphics[width=\columnwidth]{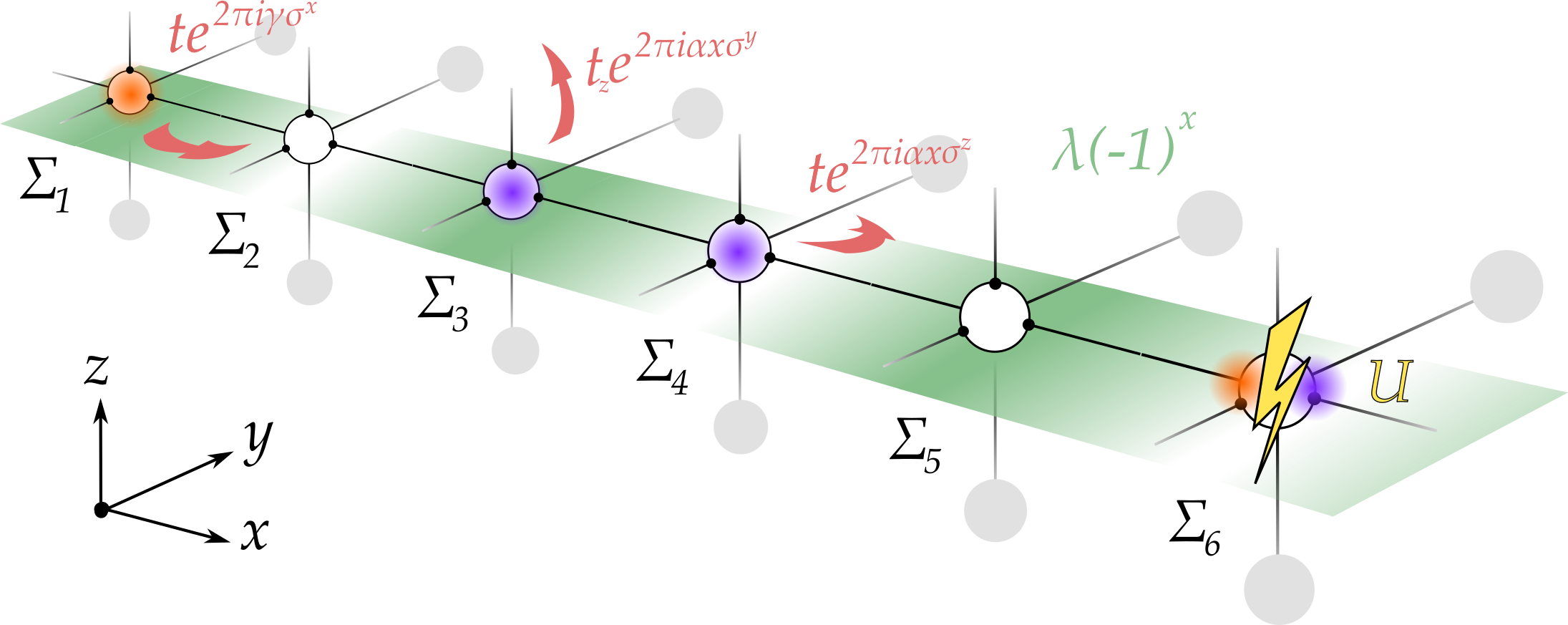}
\caption{Schematic of Hamiltonian \eqref{hamiltonian}. The parameters of the noninteracting system are described in the text. Hubbard interactions of strength $U$ are treated using dynamical mean-field theory leading to local selfenergies $\Sigma_i$ in the unit cell.}
\label{schem}
\end{figure}

In contrast to real materials, cold atomic gases allow to experimentally rebuild model Hamiltonians such as - in the context of topological states - the celebrated Hofstadter \cite{Aidelsburger2013,Miyake2013} and Haldane \cite{Jotzu2014,Flaschner2016} 2d models. 3d topological states, however, are still on their way to be experimentally accessible. In theory there are different generalizations of the Hofstadter model \cite{Kimura2014,Li2015,Zhang2017}.
Here, we study the TRI Hofstadter model \cite{Goldman2010} generalized to 3d \cite{Scheurer2015} with Hubbard interactions between two fermionic spin components. We find besides WTI and STI a NLSM in the phase diagram. In order to characterize these topologically nontrivial phases, we generalize three topological invariants to 3d, TRI, and interacting systems. Calculating surface states confirms the bulk-boundary correspondence.

The structure of the manuscript is as follows: in Sec.~\ref{mod} we introduce the 3d TRI Hofstadter-Hubbard model, in Sec.~\ref{nonInt} we discuss all quantum phases of the noninteracting system. In Secs.~\ref{wil} and \ref{loc} we put increased emphasis on the Wilson loop method and the local $\mathbb{Z}_2$ marker, respectively. We continue by discussing the interacting phases within dynamical mean-field theory in Sec.~\ref{int} as well as the corresponding surface states in Sec.~\ref{surf}. Section~\ref{con} concludes the manuscript.

\section{Model}
\label{mod}
The Hamiltonian of the 3d TRI Hofstadter-Hubbard model reads \cite{Scheurer2015}

\begin{equation}
\begin{split}
\hat{H} = \sum_{\bm{j}}\Bigg[\sum_{\mu=x,y,z}&(-t_\mu)\left(\hat{\bm{c}}_{\bm{j}+\bm{\mu}}^\dag e^{2\pi i \bm{\theta}_\mu}\hat{\bm{c}}_{\bm{j}}+\text{h.c.}\right)\\
+&(-1)^x\lambda\hat{\bm{c}}_{\bm{j}}^\dag\hat{\bm{c}}_{\bm{j}} + U\hat{c}_{\uparrow\bm{j}}^\dag\hat{c}_{\uparrow\bm{j}}\hat{c}_{\downarrow\bm{j}}^\dag\hat{c}_{\downarrow\bm{j}}\Bigg],
\end{split}
\label{hamiltonian}
\end{equation}
where $\hat{\bm{c}}_{\bm{j}}=(\hat{c}_{\uparrow\bm{j}},\hat{c}_{\downarrow\bm{j}})^T$ is the spin-1/2 fermionic annihilation spinor, $\bm{j}=(x,y,z)$ is a 3d lattice vector, $t_\mu$ is the hopping energy and $\bm{\mu}$ the unit vector in $\mu$ direction where we set $t_x=t_y=1$, $\bm{\theta}=(\gamma\sigma^x,\alpha x\sigma^z, \alpha x\sigma^y)$ is a vector of generalized matrix Peierls phases with $\gamma$ being  the spin-mixing amplitude, $\alpha$ the flux, and $\sigma^i$ the $i$th Pauli matrix. Moreover, $\lambda$ is the staggered potential amplitude in $x$ direction, and $U$ is the Hubbard interaction energy. The Hamiltonian is schematically depicted in Fig.~\ref{schem}.

\section{Noninteracting phases}
\label{nonInt}
\begin{figure}
\includegraphics[width=\columnwidth]{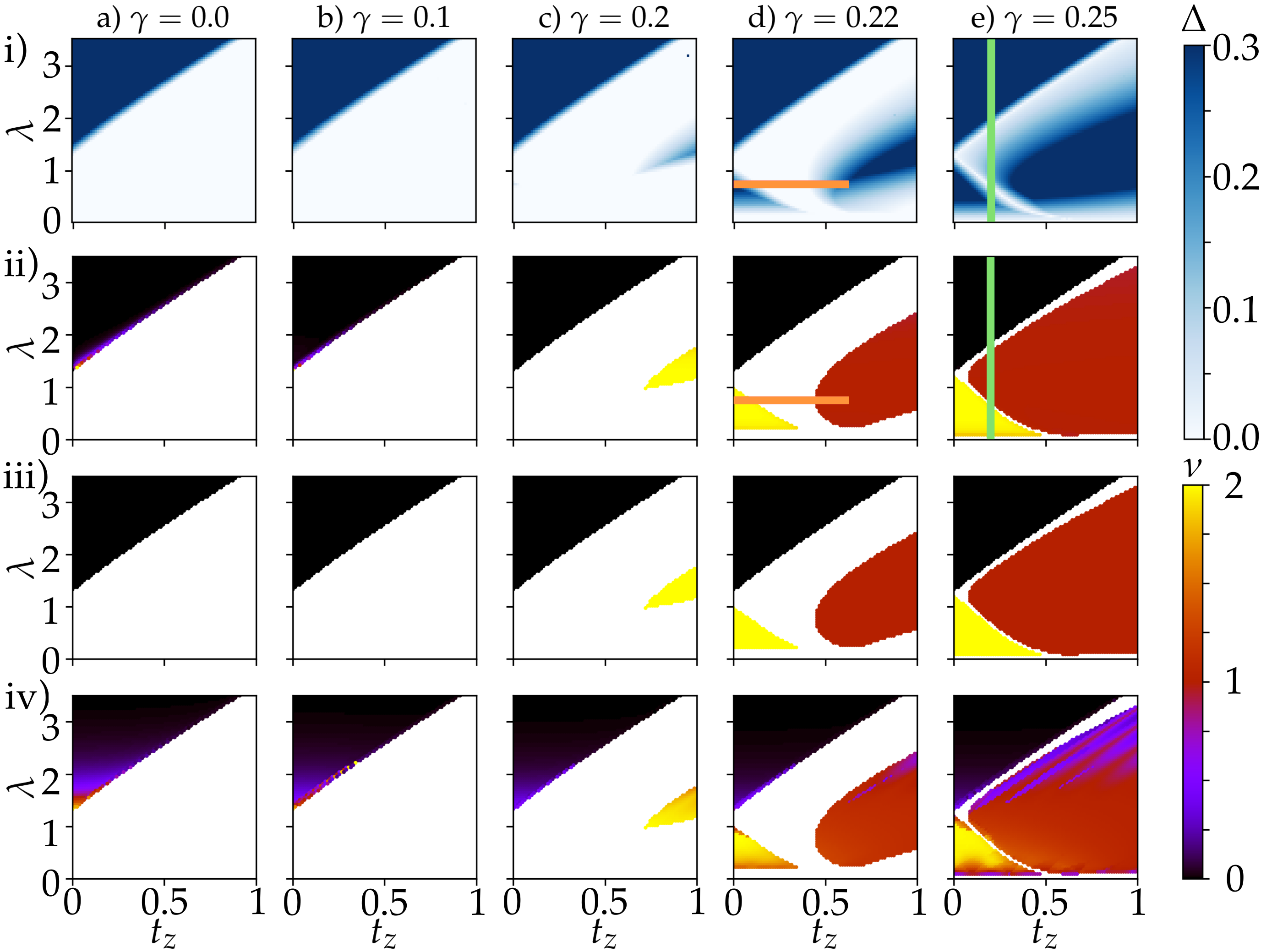}
\caption{Noninteracting, $U=0$, phase diagrams of the system described by Eq.~\eqref{hamiltonian}: i) band gap $\Delta$ and topological invariant $\nu$ defined in Eq.~\eqref{invariant} for the 3d topological insulators obtained by: ii) twisted boundary conditions, iii) Wilson loops, and iv) local $\mathbb{Z}_2$ marker for $\alpha=1/6$.}
\label{noninteracting}
\end{figure}

Before studying the interacting system, let us first understand the noninteracting case, i.e., $U=0$. We show the gap of the half-filled system \eqref{hamiltonian} in Fig.~\ref{noninteracting}i) as a function of $t_z$ and $\lambda$ for different values of $\gamma$ and $\alpha=1/6$. For $\gamma=0$ and 0.1 we find a gapped phase only for large $\lambda>1.75$ depending on the value of $t_z$. For $0.2<\gamma<0.25$ gapped phases for smaller values of $\lambda$ emerge which we will characterize by their topological invariants. The 3d topological invariants of the present system, being a stack of coupled QSH layers, can be simplified through the following invariant \cite{Scheurer2015}:
\begin{equation}
\nu = Z(0) + Z(\pi),
\label{invariant}
\end{equation}
where $Z(k_z)$ denotes the 2d $\mathbb{Z}_2$ topological index at fixed $k_z$. $\nu$ can thus assume the three values 0,1, or 2. Where 0 represents a trivial band insulator (BI) (0;0,0,0) since both 2d invariants are zero. If $\nu$ is 1, we find an STI (1;0,0,0). If both of them assume the value 1 Eq.~\eqref{invariant} assumes 2 which corresponds to the WTI with invariants (0;0,0,1). Note that we find only this particular WTI due to the chosen anisotropy $t_x=t_y\neq t_z$. From Eq.~\eqref{invariant} we understand that the 3d invariant only requires the computation of 2d $\mathbb{Z}_2$ numbers \cite{Fu2007,Moore2007,Roy2009}. In the following, we will develop and apply three different methods in order to compute the 2d $\mathbb{Z}_2$ indices for $k_z=0$ and $\pi$. We thus directly obtain the invariant \eqref{invariant}. The first method is the generalization of Fukui's method \cite{Fukui2005} to TRI systems using twisted boundary conditions (TBC) \cite{Sheng2006,Fukui2007,Kumar2016}. We first Fourier transform Eq.~\eqref{hamiltonian} for the $z$ direction. For the $x$ and $y$ direction we apply spin-dependent TBC, i.e., $\hat{\bm{c}}_{x+N_x,y,k_z}=\hat{\bm{c}}_{x,y,k_z}e^{i\vartheta_x}$ and $\hat{\bm{c}}_{x,y+N_y,k_z}=\hat{\bm{c}}_{x,y,k_z}e^{i\vartheta_y\sigma^z}$ where $N_x\times N_y$ is the size of the 2d system. Note that the spin dependence $\sigma^z$ appears only once, however, in which direction is a freedom of the gauge. After introducing the twist angles $\vartheta_x$ and $\vartheta_y$, Fukui's method is applied in $(\vartheta_x,\vartheta_y)$ space. This yields the $\mathbb{Z}_2$ invariant $Z(k_z)$ with parameter $k_z$. We show $\nu$ in Fig.~\ref{noninteracting}ii) obtained by the TBC method if the gap i) is finite. 

We find an STI phase, shown in red, as well as a WTI, shown in yellow. We also observe in Fig.~\ref{noninteracting}e) that for maximal spin mixing $\gamma=0.25$ there are gapless transition lines between the topological insulator phases. For \mbox{$\gamma=0.22$} as shown in Fig.~\ref{noninteracting}d) these transition lines extend to gapless regions which we discuss further below.

\section{Wilson loop}
\label{wil}
The second method to compute $\nu$ is the Wilson loop technique \cite{Yu2011,Grusdt2014} which is an extension of the Zak phase to multi-band systems. We present the Wilson loop technique in the 3d case and provide details for the numerical computation in the following.

We Fourier transform the Hamiltonian, defined in Eq.~\eqref{hamiltonian}, for all three spatial dimensions. For $\alpha=1/6$ the resulting $\bm{k}$-dependent Hamiltonian matrix has Hilbert space dimension 12 where the spin as well as the position $x$ within the unit cell are treated as internal degrees of freedom:

\begin{equation}
\begin{split}
\mathcal{H}(\bm{k})&=
\begin{pmatrix} 
O(1) & T&&&&T^\dag e^{ik_x}\\
T^\dag & O(2)&T\\
&T^\dag & O(3)&T\\
&&T^\dag & O(4)&T\\
&&&T^\dag & O(5)&T\\
Te^{-ik_x}&&&&T^\dag & O(6)\\
\end{pmatrix}\\
\ \\
\ \\
T&=te^{2\pi i\gamma\sigma^x}
\end{split}
\end{equation}
\begin{align*}
\begin{split}
O(x)=
&-2t\cos(k_y)\cos(2\pi \alpha x)\mathds{1}\\
&-2t\sin(k_y)\sin(2\pi \alpha x)\sigma^z\\
&-2t_z\cos(k_z)\cos(2\pi \alpha x)\mathds{1}\\
&-2t_z\sin(k_z)\sin(2\pi \alpha x)\sigma^y
+\lambda(-1)^x\mathds{1}.\\
\end{split}
\end{align*}
We define the time-reversal-invariant, gauge-independent multiband formulation \cite{Yu2011} of the discretized Wilson loop
\begin{equation}
D(C_{\bm{k}})=\prod_{\bm{k}_j\in C_{\bm{k}}}F_j,
\quad\text{with}\quad
F_j^{mn}=\langle u_m(\bm{k}_j) | u_n(\bm{k}_{j+1})\rangle,
\label{wilson}
\end{equation}
with $|u_n(\bm{k}_j)\rangle$ being the cell-periodic part of Bloch state of the $n$th band and $\bm{k}_j$ are discretized values of the closed contour $C_{\bm{k}}$ in the BZ. If we set $k_z=0,\pi$ and choose $C_{\bm{k}}$ to go along $k_x$, we find Eq.~\eqref{wilson} to be a parametric function of $k_y$ only. The eigenvalues of $D(k_y)$ are $\lambda_m(k_y)$. Their phases $\theta_m(k_y)=\mathrm{Im}\log\lambda_m(k_y)$ will perform trajectories on a cylinder which we define through $(k_y,\theta_m)\in[0,\pi]\times[0,2\pi]$. Here, $\theta_m$ is the periodic part of the cylinder.
At the ends of the cylinder, i.e., $k_y=0$ and $\pi$, the $\theta_m$ will be degenerate in pairs  due to time-reversal symmetry.
By tuning $k_y$ from $0$ to $\pi$ these pairs will split and the $\theta_m$ may wind around the cylinder. At $k_y=\pi$ the $\theta_m$ reconnect again in pairs. This integer valued winding number around the cylinder is directly connected to the time-reversal polarization \cite{Fu2006} and corresponds to the Z$_2$ number. 

\begin{figure}
\includegraphics[width=\columnwidth]{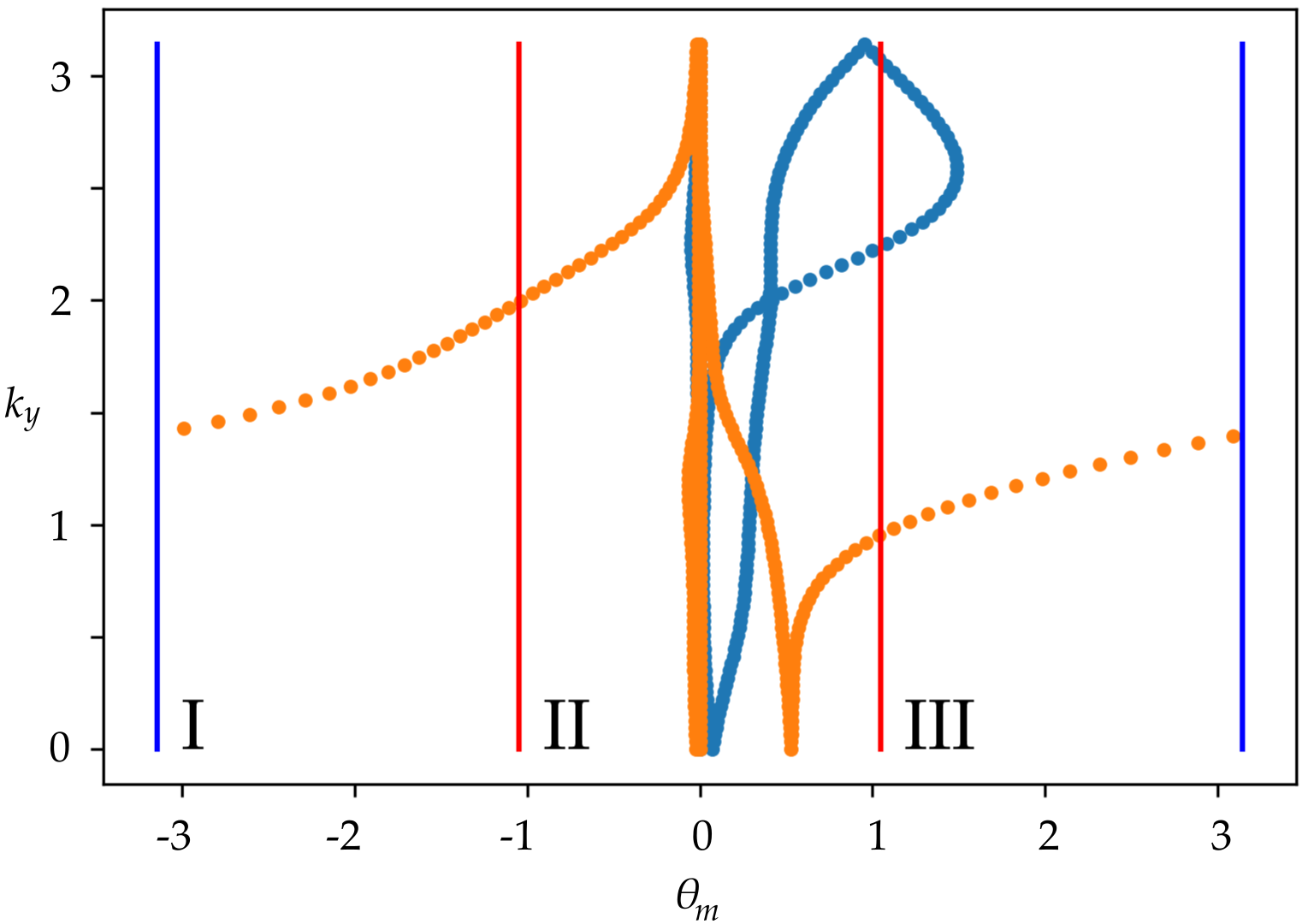}
\caption{Example for the numerical calculation of the Wilson loops. The parameters are set to $\gamma=0.22$, $t_z=0.6$ and $\lambda=0.75$. The blue data points correspond to $k_z=0$ and the orange data points to $k_z=\pi$.}
\label{cylinder}
\end{figure}

Numerically, we find this winding number by dividing the cylinder into three regions: I where $0<\theta_m<2\pi/3$, II where $2\pi/3<\theta_m<4\pi/3$, and III where $4\pi/3<\theta_m<2\pi$. The winding is depicted in Fig.~\ref{cylinder} for a trivial (blue) and a nontrivial instance (orange). We sample a sufficient set of values of $k_y$ and count the number $n_i$ of $\theta_m$ values being in the region $i$, with $i$=I,II,III. This yields the data $(n_1,n_2,n_3)$ as a function of $k_y$. We then compute the change $\Delta n_i$ of $n_i$ with respect to $\Delta k_y$. From this data, we only keep the ones where $(\Delta n_1,\Delta n_2,\Delta n_3)$ follow some permutation of $-1,0,1$. Data, where $(\Delta n_1,\Delta n_2,\Delta n_3)$ all are zero, do not carry information and data where a 2 appears could be removed by increased sampling of $k_y$ and can thus be safely omitted. Finally, to each of the remaining data points a chirality can be assigned by means of the Levi-Civita tensor. Summing these chiralities yields a nontrivial $\mathbb{Z}_2$ number for odd and a trivial one for even values of the sum of the chiralities. 
The results are shown in Fig.~\ref{noninteracting}iii) and they agree exactly with the method using TBC in Fig.~\ref{noninteracting}ii).

\section{Local $\mathbb{Z}_2$ marker}
\label{loc}
We now turn to the generalization of the local Chern marker \cite{Bianco2011} to the TRI case, first in 2d. A $\mathbb{Z}_2$ generalization to Kitaev's real-space formulation of the Chern number \cite{Kitaev2006} has recently appeared in Ref.~\cite{Li2019}. Here, we introduce the spin-projected version of the local Chern marker

\begin{equation}
C_{\mu\nu}(x,y)=\langle x,y|\mathcal{P}_\mu\hat{P}\hat{x}\hat{P}\hat{y}\hat{P}\mathcal{P}_\nu |x,y\rangle,
\end{equation}
where $\mathcal{P}_\mu$ is the projector onto the states of band $\mu=$I,II in the \{I,II\} eigenbasis of time-reversed partners \cite{Fu2006}, $\hat{P}$ is the  projector onto the occupied eigenstates of the Hamiltonian, and $|x,y\rangle$ is the eigenstate of the position operator in 2d. The eigenvalues of the 2$\times$2 matrix $C_{\mu\nu}(x,y)$ correspond to time-reversed partners similar to the partial polarizations of the time-reversed partners in Ref.~\cite{Fu2006}, however, now defined in real space. The first eigenvalue thus resembles exactly the 2d local $\mathbb{Z}_2$ marker. Since the eigenvalues are independent of the basis in which $C_{\mu\nu}(x,y)$ is represented, we can also use the spin basis \{$\uparrow,\downarrow$\} such that we do not have to find the \{I,II\} basis. By Fourier transforming only the $z$ coordinate of Eq.~\eqref{hamiltonian} and fixing the value of $k_z=0,\pi$ we can generalize the 2d local $\mathbb{Z}_2$ marker to a 3d local $\mathbb{Z}_2$ marker. 

The bulk value is presented in Fig.~\ref{noninteracting}iv) showing approximately the same behaviour as the aforementioned methods in Fig.~\ref{noninteracting}ii) and iii) computed on a 30$\times$30 lattice. However, the local $\mathbb{Z}_2$ marker suffers from finite size effects when the gap is small \cite{Irsigler2019b}. This can be observed, e.g., in Fig.~\ref{noninteracting}iv)e) for large $t_z$ and $\lambda$ where the local $\mathbb{Z}_2$ marker is not quantized due to the finite system. If the gap is sufficiently large, however, the local $\mathbb{Z}_2$ marker is well quantized.

\begin{figure}
\includegraphics[width=\columnwidth]{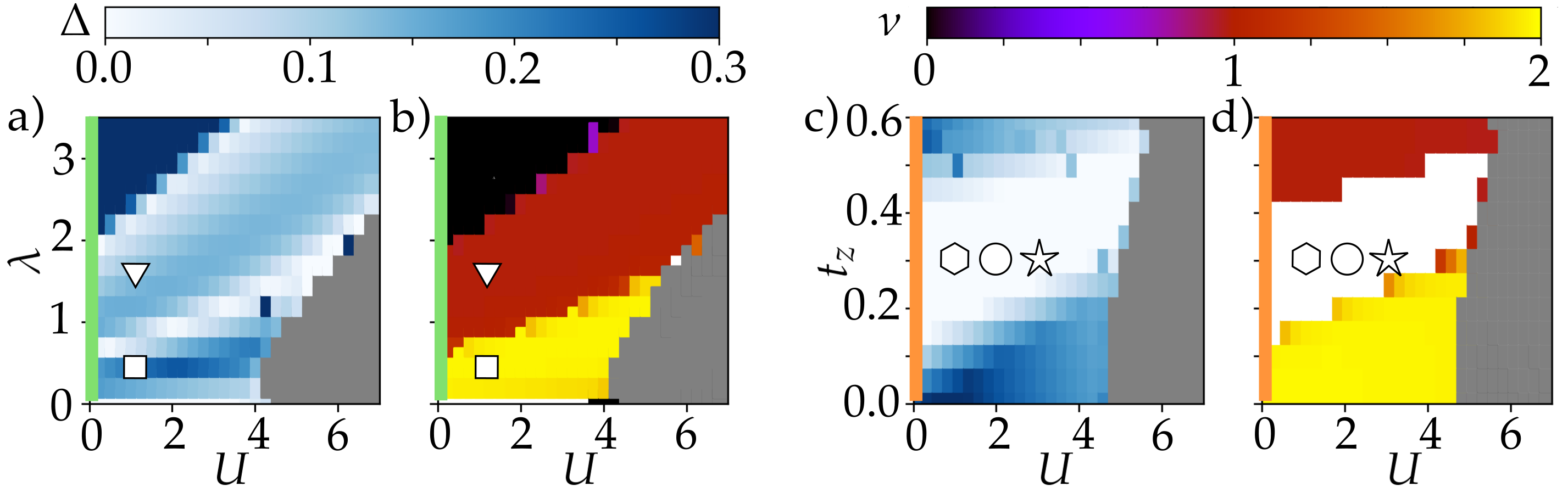}
\caption{Interacting phase diagrams obtained from DMFT and the topological Hamiltonian Eq.~\eqref{topHam}. a) shows the band gap of the topological Hamiltonian $\Delta$ and b) the topological invariant $\nu$, Eq. \eqref{invariant}, as functions of the interaction strength $U$ and the staggered potential $\lambda$ for $\gamma=0.25$ and $t_z=0.2$. The green lines correspond to the green lines in Fig.~\ref{noninteracting}d)i) and ii). c) shows $\Delta$ and d) $\nu$ as functions of $U$ and $t_z$ for $\gamma=0.22$ and $\lambda=0.75$. The orange lines correspond to the orange lines in Fig.~\ref{noninteracting}e)i) and ii). The gray regions denote DMFT solutions which break the lattice symmetry. The white symbols correspond to the parameter sets used in Fig.~\ref{surfStates}.}
\label{interacting}
\end{figure}

\section{Interacting phases}
\label{int}
We study interaction effects by applying dynamical mean-field theory (DMFT) \cite{Georges1996} which neglects nonlocal fluctuations but covers all local fluctuations. Since the unit cell of the system \eqref{hamiltonian} contains six lattice sites if there is no spontaneous symmetry breaking, we make use of the real-space version of DMFT \cite{Okamoto2004,Helmes2008,Snoek2008}. Here, the many-body problem of the full  lattice with $N$ sites is mapped onto $N$ single-site quantum impurity problems, where each impurity problem interacts with a self-consistent, noninteracting bath. This approach nonperturbatively describes local quantum dynamics, in contrast to static mean-field theory. After solving the single-impurity problem for each site, for which we use exact diagonalization with four bath sites, the selfenergy $\Sigma_{\bm{j}}^{\sigma\sigma'}(\omega)$ for each lattice site $\bm{j}$ and frequency $\omega$ is obtained. Here, $\sigma,\sigma'$ are spin degrees of freedom. Using the Dyson equation, these are used to construct a new lattice Green's function and this procedure is repeated until self-consistency.
In 2d DMFT has provided a successful description of topological systems for many aspects \cite{Cocks2012,Orth2013,Vanhala2016,Kumar2016,Amaricci2017,Irsigler2019,Irsigler2019a,Gebert2019}.
Ref.~\cite{Mertz2019} has shown that nonlocal contributions are small already in 2d. We therefore expect even more accurate results of DMFT in 3d.

To calculate topological invariants for the interacting system we follow the topological Hamiltonian approach \cite{Wang2012}. The idea here is, that if the Green's function can be smoothly deformed to a noninteracting Green's function, i.e., no poles or zeros occur, the topological properties do not change. This holds since topological phase transitions come along with a divergence or a zero \cite{Gurarie2011,Zheng2017} of the Green's function. In this way, one can construct an effective, noninteracting Hamiltonian $H_T=-G^{-1}(\omega=0)$ which is used to compute topological invariants. In combination with the local selfenergy $\Sigma^{\sigma\sigma'}_{\bm{j}}(\omega)$ from DMFT its matrix form reads:

\begin{equation}
\left[H_T\right]^{\sigma\sigma'}_{\bm{j}\bm{j'}} = \left[H_0\right]^{\sigma\sigma'}_{\bm{j}\bm{j'}}+\Sigma^{\sigma\sigma'}_{\bm{j}}(\omega=0)\delta_{\bm{j}\bm{j'}},
\label{topHam}
\end{equation}
where $H_0$ denotes the noninteracting part of the Hamiltonian. We show the gap $\Delta$ as well as the topological invariant \eqref{invariant} of the topological Hamiltonian \eqref{topHam} in Figs.~\ref{interacting}a) and b), respectively, as functions of $\lambda$ and $U$ for $\gamma=0.25$ and $t_z=0.2$. The green lines for $U=0$ correspond to the green lines in Figs.~\ref{noninteracting}e)i) and ii), respectively. The grey regions correspond to DMFT results where the lattice symmetry is spontaneously broken. 

We first focus on the symmetric phases. We observe that the gap closing lines in Fig.~\ref{interacting}a) coincide with the topological phase transition lines in b) as expected. Furthermore, we find stabilization of the STI and the WTI phases against $\lambda$ through $U$. For small $U$, Hubbard interactions effectively renormalize $\lambda$, which extends the topological phases in the phase diagram. This is the 3d analogue of the interaction-induced topological phase transition in 2d \cite{Kumar2016,Zheng2018} and can be understood through the competition between staggered potential and interactions.

In Figs.~\ref{interacting}c) and d) we present $\Delta$ and $\nu$ as functions of $t_z$ and $U$. The orange lines for $U=0$ correspond to the orange lines in Figs.~\ref{noninteracting}d)i) and ii), respectively, for $\gamma=0.22$ and $\lambda=0.75$.  As in the previous phase diagram, we observe again stabilization of the topological phases through interactions.

We compare our results to a DMFT study of a four-band model including a Hund's coupling term \cite{Amaricci2016}. We find qualitative agreement between the phase diagram in Fig.~\ref{interacting}d) and the one in Ref.~\cite[Fig.~3]{Amaricci2016} even though the latter corresponds to a finite Hund's coupling which we do not include here. This is because the Hund's coupling effectively reduces the interorbital interactions and thus makes the Hubbard term the dominant interaction term. In contrast to Ref.~\cite{Amaricci2016}, we do not find the (1;1,1,1) phase.

\begin{figure}[t]
\includegraphics[width=1\columnwidth]{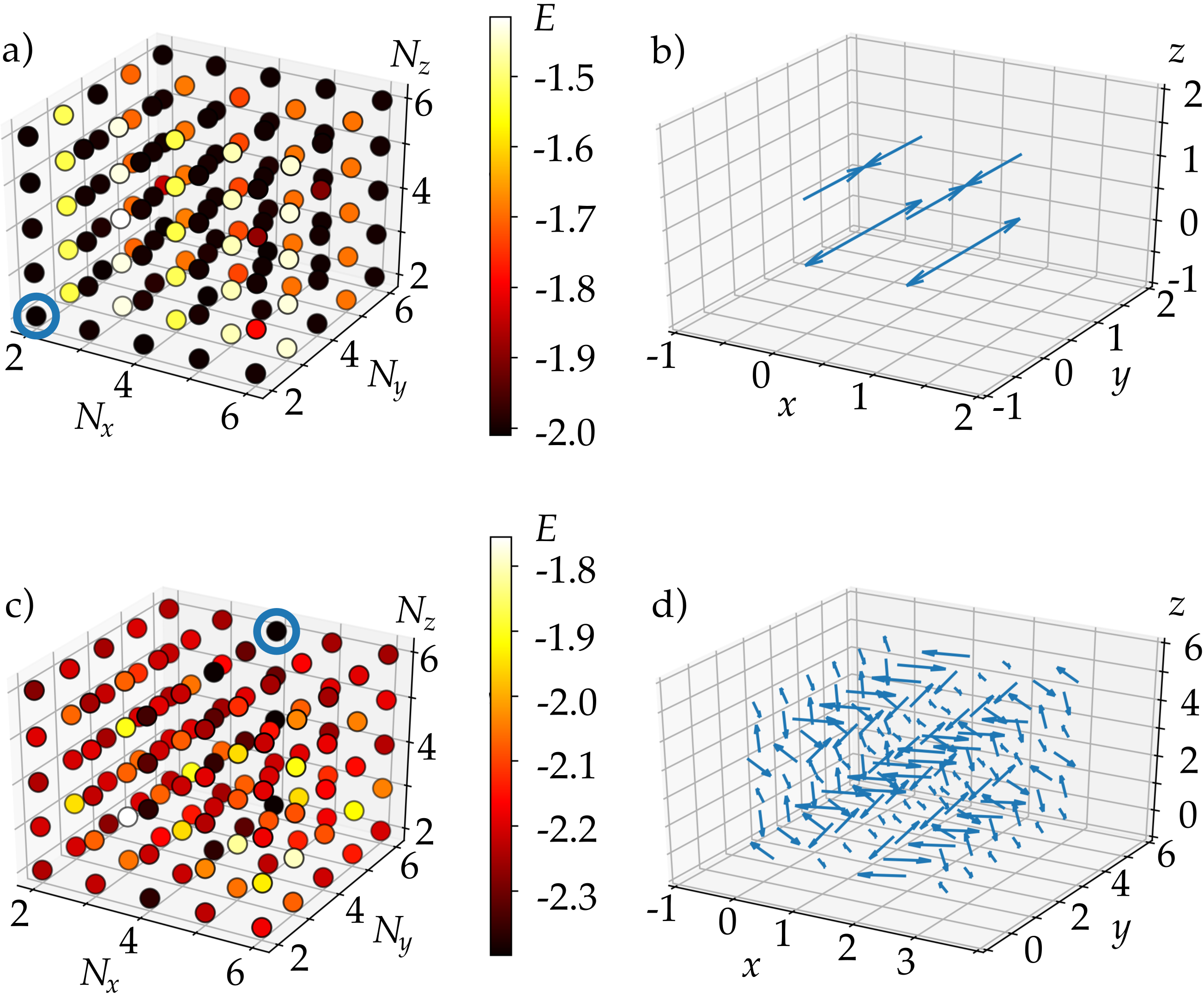}
\caption{Classical Monte Carlo results for the groundstate energy of Eq.~\eqref{spin} for a) $\gamma=0.25$ and $t_z=0.2$ as well as for c) $\gamma=0.22$ and $t_z=1$ as function of the size of the unit cell $N_x\times N_y\times N_z$. b) and d) show the spin state of the unit cell with the smallest energy marked by a blue circle in a) and c), respectively. }
\label{monte}
\end{figure}

It is a priori not clear what would be the unit cell of a possible spontaneous-symmetry-broken phase as a result of the nontrivial exchange couplings between neighboring spins due to the Peierls phases in Eq.~\eqref{hamiltonian}. The effective spin Hamiltonian \cite{Cocks2012} for the 3d system reads:
\begin{equation}
\begin{split}
\hat{H}_\text{spin} = \sum_{\bm{j}}&\sum_{\mu,\nu,\rho\text{ cyclic}}\frac{t_\mu^2}{U}\left\{\hat{S}^\mu_{\bm{j}}\hat{S}^\mu_{\bm{j}+\bm{\mu}}\right.\\
&\left.+\cos(4\pi\theta_\mu)\left[\hat{S}^\nu_{\bm{j}}\hat{S}^\nu_{\bm{j}+\bm{\mu}}+\hat{S}^\rho_{\bm{j}}\hat{S}^\rho_{\bm{j}+\bm{\mu}}\right]\right.\\
&\left.+\sin(4\pi\theta_\mu)\left[\hat{S}^\nu_{\bm{j}}\hat{S}^\rho_{\bm{j}+\bm{\mu}}-\hat{S}^\rho_{\bm{j}}\hat{S}^\nu_{\bm{j}+\bm{\mu}}\right]\right\},
\end{split}
\label{spin}
\end{equation}
where we defined the spin operator $\hat{S}^\mu_{\bm{j}}=\hat{\bm{c}}_{\bm{j}}^\dag\sigma^\mu\hat{\bm{c}}_{\bm{j}}$ and $\theta=(\gamma,\alpha x, \alpha x)$. In the spin population balanced, 2d case \cite{Orth2013,Irsigler2019a} one can argue that the spins will always order antiferromagnetically in $y$ direction. However, in the 3d case we cannot make this argument and the unit cell might in fact be very large. Results of a classical Monte Carlo study to find the classical ground state $E$ of Eq.~\eqref{spin} are shown in Fig.~\ref{monte} for unit cells up to $N_x\times N_y\times N_z=6^3$ lattice sites. Examplarily, for $\gamma=0.25$ and $t_z=0.2$ in a) the unit cell with the smallest energy is found to have dimensions $2\times2\times2$ and marked by a blue circle. This state is shown in b) and corresponds to ferromagnetic ordering in $x$ and anti-ferromagnetic ordering in $y$ and $z$ direction similar to the 2d collinear order in Ref.~\cite[Fig.~6]{Orth2013}. For $\gamma=0.22$ and $t_z=1$ the unit cell with smallest energy is found with dimensions $4\times6\times6$ and is marked by a blue circle in Fig.~\ref{monte}c). The corresponding spin state is shown in d) and corresponds to a spiral order in all the spatial directions. We conclude that within the symmetry-broken phase there must be phase transitions between ferro/anti-ferromagnetic and spiral orders, which depend on the parameters $\gamma$ and $t_z$ and can lead to complex magnetic orders.

\section{Surface states}
\label{surf}

\begin{figure}[t]
\includegraphics[width=\columnwidth]{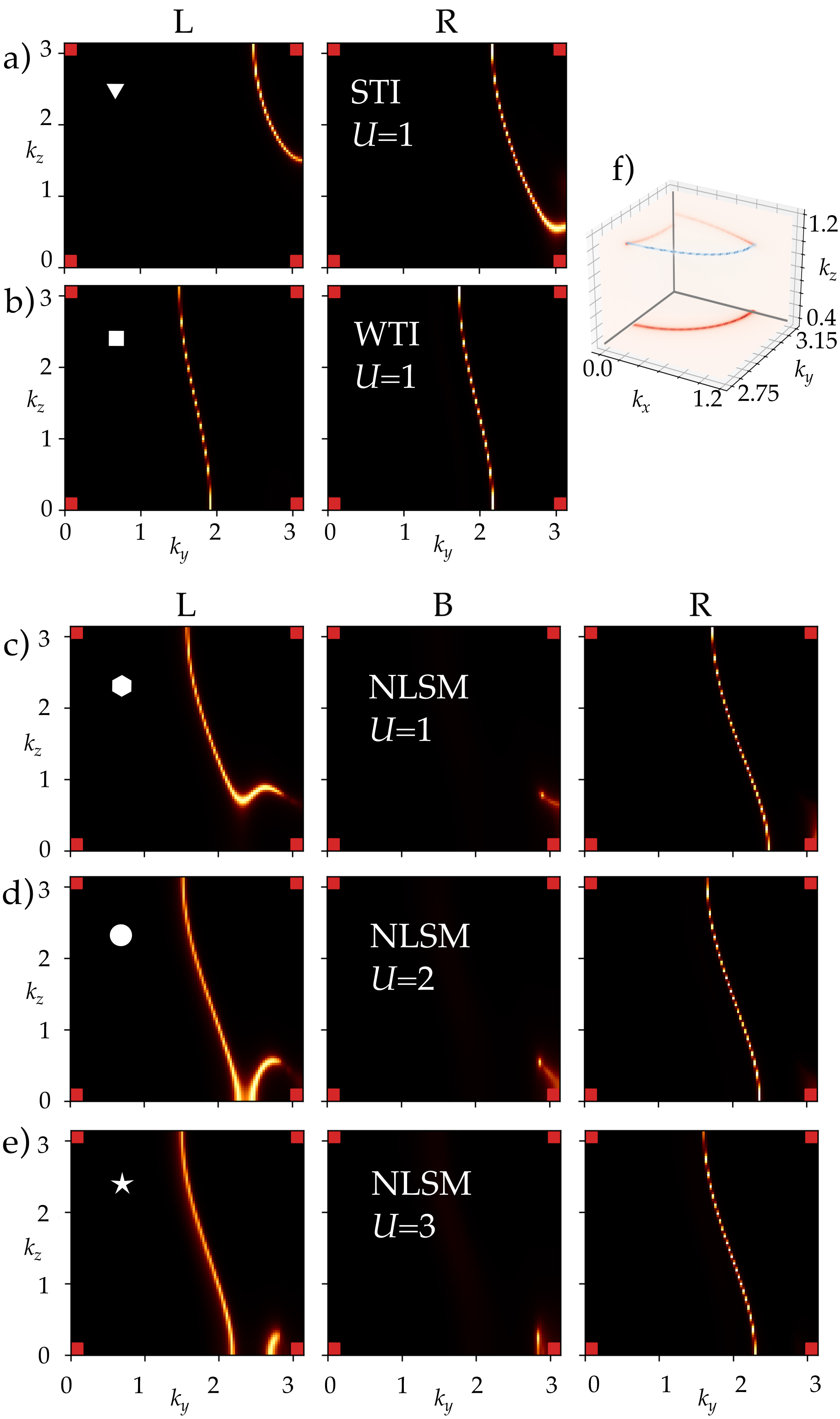}
\caption{Surface states of different 3d, interacting, topologically nontrivial phases: a) strong topological insulator, b) weak topological insulator, and c)-e) nodal-line semimetals. The white symbols correspond to the parameter sets marked by the symbols in Fig.~\ref{interacting}.  See definitions of L, B, and R below Eq.~\eqref{spectralDensity}. The bulk nodal line in the full 3d BZ is shown in f) in blue corresponding to the projected nodal line in c)B; projections onto the $k_i$-$k_j$ planes for $i,j=x,y,z$ are shown in red.}
\label{surfStates}
\end{figure}

We now study the surface states of the present system. To this end we put the system on a 3d cylinder geometry, i.e., $k_y,k_z$ are good quantum numbers but in $x$ direction we now apply open boundary conditions. For this geometry, we can define the single-particle Green's function:

\begin{equation}
G^{\sigma\sigma'}_{xx'}(\omega,k_y,k_z)=
\left[\left\{\omega-\Sigma(\omega)-H_0(k_y,k_z)\right\}^{-1}\right]^{\sigma\sigma'}_{xx'},
\end{equation}
where $x,x'$ are the spatial degrees of freedom in $x$ direction. The spectral density of a spatial region $X$ is defined as

\begin{equation}
\rho_X(\omega,k_x,k_y)=-\frac{1}{\pi}\sum_{\sigma,x\in X}\text{Im}G^{\sigma\sigma}_{xx}(\omega,k_y,k_z).
\label{spectralDensity}
\end{equation}
We show the surface states of a system with 60 sites in $x$ direction by plotting $\rho_X(\omega=0,k_y,k_z)$ in Fig.~\ref{surfStates} for the left surface L where $1\leq x\leq 3$, the bulk B where $25\leq x\leq36$, and the right surface R where $58\leq x\leq60$. We plot only one quarter of the surface BZ since the results are mirror symmetric at the two lines $k_y=0$ and $k_z=0$. The red dots denote TRI momenta. The parameters are chosen according to the white symbols in Fig.~\ref{interacting}. For $U=1$, Fig.~\ref{surfStates}a) shows the Fermi surface enclosing only one TRI momentum which corresponds to the surface state of an STI. Figure~\ref{surfStates}b) shows the Fermi surface crossing the BZ almost parallel to the $k_z$ axis. Thus it encloses two TRI momenta which corresponds to the surface state of a WTI. We do not show the results for the bulk because it is gapped.

Let us now turn to the case of the NLSM. Figures~\ref{surfStates}c) to e) show the surface states as well as the $x$ projection of the bulk state for different $U$. First, we notice that the two surfaces are asymmetric. This arises from the broken parity symmetry since upon the transformation $\lambda\rightarrow-\lambda$ the spectral density behavior of the two surfaces is exchanged. The right surface shows a state corresponding to a WTI, whereas the left surface rather shows one corresponding to the STI. However, the surface state of the left surface does not fully enclose the TRI momentum but rather stops within the BZ, reminiscent of a Fermi arc. The missing part is recovered in the bulk and identifies as a nodal line. It is shown for the full 3d BZ in Fig.~\ref{surfStates}f).

In order to justify that the nodal line is not an accidental band touching but is topologically protected we compute the Berry phase on a closed path in the 3d BZ linking the nodal line. To this end, we make use of the multiband formulation of Ref.~\cite{Yu2011} to find the Berry phase $\text{Im}\text{Tr}\log\prod_jF_j$, where $F_j^{mn}=\langle u_m(\bm{k}_j) | u_n(\bm{k}_{j+1})\rangle$ and $|u_n(\bm{k}_j)\rangle$ is the cell-periodic part of Bloch state of the $n$th band. If the path $\bm{k}_j$ is (is not) linked with the nodal line the Berry phase yields $\pi$ (0). This shows that the nodal line is topologically protected and thus cannot gap out. On the other hand, when computing the Berry curvature on a 2d box surface enclosing the nodal line we find a vanishing Chern number. Thus the nodal line does not carry a topological charge.

Refs.~\cite{Sur2016,Roy2017,Kang2019} studied the interacting NLSM with renormalization group calculation, static mean-field, and cluster perturbation theory, respectively. Since the density of states of a NLSM vanishes at the band touching the nodal line is robust against interactions and will only gap out for strong interactions. On the other hand, interactions can change the size and the shape of the nodal line.

The surface states of NLSMs have attracted a lot of attention. If they are drumhead states, they constitute flat bands with a diverging density of states which is localized at the surfaces of the system. Ref.~\cite{Liu2017} have studied emergent surface antiferromagnetic order in this context already for small critical interaction strength, which is, however, increased by spin-orbit coupling. We do not observe any symmetry breaking in our parameter range and do not find this surface magnetism. We attribute this to the strong spin-orbit coupling in our system which curves the flat bands and thus decreases the surface density of states. A recent study  found an inversion of the Berry curvature of one $\bm{K}$ point driven by spin-orbit coupling and two-particle interactions in the Haldane model, which leads to a surface Chern insulator \cite{Chen2019}. Bulk antiferromagnetism in NLSMs was investigated in Ref.~\cite{Wang2017}.

From the experimental point of view, the TRI Hofstadter Hamiltonian has been realized in 2d using laser-assisted tunneling \cite{Aidelsburger2013}. Theoretically, this approach has been generalized to 3d \cite{Zhang2016,Zhang2017,Chen2017}. A generic way to implement spin-orbit coupling proposed in Ref.~\cite{Grusdt2017} might be generalizable to three dimensions. For detection, Bloch-Zener-St\"uckelberg interferometry \cite{Zhang2016}, anomalous velocity measurement, or state tomography \cite{Zhang2017} for bulk states as well as Bragg spectroscopy for surface states \cite{Chen2017} have been proposed. Also a 3d version of a topological interface \cite{Irsigler2019} could be used for the detection of the surface states. The here introduced local $\mathbb{Z}_2$ marker could be used to distinguish the topological phase at the interface. Very recently, a NLSM has been realized in a fermionic cold atom experiment with $^{173}$Yb atoms by mapping the $k_z$ component to a Zeeman field and reading out 2d layers for each value of $k_z$ \cite{Song2019}.

\section{Conclusion}
\label{con}
We develop three numerical techniques for the characterization of three-dimensional topological states of matter. Based on twisted boundary conditions, Wilson loops, and the local topological marker, these techniques can be used to compute weak and strong topological indices even in interacting systems. We apply these to the three-dimensional time-reversal-invariant Hofstadter-Hubbard model and find a topological nodal-line semimetal between phases of weak and strong topological insulators. Using dynamical mean-field theory we observe stabilization of the three-dimensional topological states through Hubbard interactions. The numerical methods presented here enable the study of interacting, three-dimensional topological matter in inhomogeneous systems, which will be of great interest for cold atomic implementations. Moreover, we think that our results could contribute to benchmark circuit-based quantum simulators where is has been possible to engineer artificial gauge fields as well as strong interactions \cite{Koch2010,Roushan2017,Owens2018}.

\begin{acknowledgments}
The authors acknowledge useful discussions with Mohsen Hafez-Torbati.
This work was supported by the Deutsche Forschungsgemeinschaft (DFG, German Research Foundation) under Project No. 277974659 via Research Unit FOR 2414 and Germany’s Excellence Strategy - EXC2111 - 390814868. This work was also supported by the DFG via the high performance computing center LOEWE-CSC.
\end{acknowledgments}

\bibliographystyle{apsrev4-1}
\bibliography{lib}

\end{document}